\documentclass[%
 reprint,
superscriptaddress,
nofootinbib,
 amsmath,amssymb,
 aps,
]{revtex4-2}
\usepackage{graphicx}
\usepackage{dcolumn}
\usepackage{bm}
\usepackage{hyperref}
\begin{document}

\preprint{APS/123-QED}
\title{Evolution of perturbation and power spectrum in a two-component ultralight axionic universe}
\author{Yi-Hsiung Hsu}
 \email{r07244003@ntu.edu.tw}
 \affiliation{%
 Institute of Astrophysics, National Taiwan University, Taipei 10617, Taiwan
 }%
 \author{Tzihong Chiueh}%
 \email{chiuehth@phys.ntu.edu.tw}
 \affiliation{%
 Institute of Astrophysics, National Taiwan University, Taipei 10617, Taiwan
 }%
\affiliation{%
 Department of Physics, National Taiwan University, Taipei 10617, Taiwan
 }%
 \affiliation{%
 Center for Theoretical Physics, National Taiwan University, Taipei 10617, Taiwan
 }%
 
\date{\today}

\begin{abstract}
The evolution of cosmic perturbations in a two-component ultralight axionic universe is investigated. We present the first spectral computation of perturbations in multi-component universes. A particular case composed of light extreme axions and free massive particles offers a possibility for the formation of very high-redshift massive galaxies, which are typically required to host massive early quasars. Our computation retains the information of perturbed velocities for individual axion components, opening a new avenue for setting up initial conditions for future axion dark matter simulations.
\end{abstract}

\maketitle
\section{Introduction}
Dark matter has played an essential role in the structure formation of the universe. While cold dark matter (CDM) has long been successful in determining large scale structures, it may fail in explaining small scale features~\cite{FDM_intro1994-2}. Ultralight axion dark matter (ADM), which has its string origin, has been proposed as an alternative to solve this problem~\cite{FDM_intro1994,FDM_intro1999,FDM_intro2000-1,FDM_intro2000-2,FDM_intro2000-3,FDM_stringreview,Schive_2014,Schive_2014_prl}, and some recent reviews can be found in~\cite{FDM_review-1,FDM_intro2017,FDM_review-2}. 

While successful in accounting for the small-scale spectral deviation, the axion mass tension has posed a problem to this model. In particular, analyses from Lyman-$\alpha$ forest give a constraint on mass lower bound at $10^{-21}$eV~\cite{mass_bound2.9-21_2017,mass_bound-21_2017}. The mass has even been suggested to be larger than $10^{-20}$eV with a small window at $10^{-21}$eV in ~\cite{mass_bound-20_2019}. On the other hand, analyses on dwarf galaxies have shown that ADM favors mass around $10^{-22}$eV~\cite{mass_1.18-22_2017,mass_3-22_2019}. In addition, the superradiance of M$87^*$ has excluded the mass range $2.9\times 10^{-21}\sim4.6\times 10^{-21}$eV~\cite{superradiance}. This inconsistency in axion masses has been a major problem to this model. Though the so-called extreme axion model has been proposed to solve this particular problem~\cite{lyman-a}, multi-component universes could also offer a solution with more degrees of freedom, as shown in this work.

On the other front, a central compact star cluster has been observed in the dwarf galaxy Eridanus II~\cite{Li_2017}. In the ADM context, the star cluster should sit inside a soliton, one of the unique nonlinear features of the ADM. However, the soliton random walk is prone to destroy the observed central star cluster; although one solution has been proposed in~\cite{starcluster}, their simulation has imposed a $5$ Gyr bound on the star cluster's age. Two-component universes may provide a different scenario where the star cluster is strongly trapped by a second more compact soliton. In this regard, the age constraint could be alleviated. 

The multi-component universe has recently been a rising research area~\cite{twocomp1,Twocomponent,ADMCDMsim}. In this paper, we explore the potential of multi-component universes. We aim to analyze linear perturbations specifically for two-component universes. We focus on the mass range from $10^{-22}$eV to $10^{-20}$eV and produce their power spectra. We also compute a universe composed of $10^{-22}$eV and $10^{-23}$eV axions, which shows a feature that could generate high red-shift massive galaxies.

Ultralight axion perturbations in the radiation epoch have been analyzed in~\cite{Zhang1,Zhang2}, while those in the matter era can be found in~\cite{Woo_2009}. Our computation follows much in these papers, especially~\cite{Zhang1} and~\cite{Zhang2}, which are engaged in the evolution for complete perturbed equations (full evolution). However, the full-evolution approach is formidably time-consuming to compute due to the stiffness of the differential equations. Hence, we shall make some approximations to remove the equation stiffness and execute the computation efficiently. To smooth out the rapid mass oscillations of the Klein-Gordon equation, we adopt the Schr\"{o}dinger equation after the onset of mass oscillations and the horizon entry. For photon and baryon perturbations, they involved strong Thomson scattering at the early time, and this scattering entails very small time steps for integration. We, therefore, utilize the diffusion approximation to cure this problem. The latter approximation only works between the time of horizon entry and when strong Thomson scattering becomes feeble. After the mean free path ($l_T$) becomes comparable to the wavelength of mode $k$, we need to switch from the diffusion approximation back to the original friction coupling. 

We find these improvements increase the efficiency of computing the linear perturbations by several orders of magnitude and provides a new way to generate initial conditions at $z\leq100$ for future ADM simulations. In addition, the goodness of matching conditions bridging the full evolution and the approximations are carefully evaluated in this paper.

We organize the paper as the following. Sec.~\ref{sec:Approx} shows how the evolution of linear perturbations is computed. We compare our results with those of the full-evolution computation in the one-component universe in Sec.~\ref{sec:onecom}. Results of the two-axion universes are presented in Sec.~\ref{sec:twocom}. Finally, our conclusions are given in Sec.~\ref{sec:conclusion}.  We place the discussion of limitations in our computation in Appendix~\ref{appd}.

Throughout this paper, the chosen cosmological parameters are $\Omega_b = 0.06$, $\Omega_{DM} = 0.24$, and $H_0 = 70$ km/sec/Mpc. The equations are written in nature units; that is, speed of light $c$ and Planck constant $\hbar$ are both set to $1$. Scale factor and conformal time are represented by $a$ and $\tau$, respectively. Conformal time is related to cosmic time with $dt = ad\tau$, and the Hubble parameter is defined as $H \equiv d\ln a/d\tau$. We adopt Newtonian gauge for the perturbations.
\section{Time step Reduction Scheme}
\label{sec:Approx}
We separate the evolution of axion dark matter into three phases. The first phase is at the time when a $k$ mode is super-horizon or before the onset of mass oscillations. In this phase, we adopt the full-evolution equations, where the Klein-Gordon equation is used for the axion field. For photon and baryon perturbations, a set of frozen-in equations, which take baryons and photons to be the same component with a $3:4$ density ratio, is adopted. The second phase starts after the mode enters the horizon, mass oscillations begin, and Thomson scatterings are moderate. Schr\"{o}dinger equation for axions and diffusion approximation for baryons and photons are utilized in this phase. In the final phase, we switch from the diffusion approximation back to the friction equations when the Thomson scattering is weakened, with some modifications discussed at Sec.~\ref{sec:pb}, while keeping the Schr\"{o}dinger equation.
\subsection{Axion model}
\label{sec:am}
\subsubsection{Full evolution}
\label{sec:fea}
The chosen axion potential is $V(\Theta_i) = m_i^2 f_i^2[1-\cos(\Theta_i)]$, where $m$ is the axion mass, $f$ is the axion decay constant, $\Theta$ is the angle of background axion, and the subscript $i$ indicates the axion component index. The axion field $\Phi_i$ relates the decay constant with the background angle via $\Phi_i = f_i\Theta_i$. The governing equation for the background angle is
\begin{eqnarray}
\label{bgeom}
\Theta_i'' + 2H\Theta_i + m_i^2 a^2 \sin(\Theta_i) = 0,
\end{eqnarray}
where the primes denote $d/d\tau$. For the axion angle perturbation $\delta \Theta$, the equation of motion for the $k$ mode is 
\begin{eqnarray}
\label{apeom}
\delta \Theta_i'' + 2 H \delta \Theta_i' +&& [k^2+m_i^2 a^2 \cos(\Theta_i)] \delta \Theta_i \nonumber \\
&&= 4 \phi' \Theta_i' - 2m_i^2 a^2 \sin(\Theta_i) \phi,
\end{eqnarray}
where $\phi$ is the metric perturbation for the Newtonian gauge. The metric perturbation $\phi$ can be defined as follows:
\begin{eqnarray}
\phi = -\frac{4\pi Ga^2}{k^2}&&\left[ \vphantom{\sum_{i}}\epsilon_\gamma \Delta_\gamma + \epsilon_\nu \Delta_\nu \right. \nonumber \\
&&\left.+\epsilon_b \Delta_b+\sum_{i} \epsilon_{A, i} \Delta_{A, i} \right], 
\label{phi}
\end{eqnarray}
where $\epsilon$ denotes the background energy, subscript $\gamma$, $b$, $\nu$, and $A$ denotes photon, baryon, neutrino, and axion, respectively, and $\epsilon_{A, i} \equiv (f_i^2/a^2)\{(\Theta_i')^2/2+m_i^2[ 1-\cos(\Theta_i) ] \}$; the covariant energy densities $\Delta$ are defined as $\Delta_\gamma \equiv \delta_\gamma - 4H\theta_\gamma$, $\Delta_\nu \equiv \delta_\nu - 4H\theta_\nu$, $\Delta_b \equiv \delta_b - 3H\theta_b$, with $\delta$ being the dimensionless energy perturbation scaled to its background energy and $\theta$ being the perturbed velocity potential which relates to the velocity $v$ by $v = ik\theta$, and
\begin{eqnarray}
\Delta_{A, i}
\equiv \frac{\Theta_i' \delta\Theta_i' +m_i^2 a^2 \sin{\Theta_i} \delta\Theta_i - (\Theta_i')^2\phi}{(1/2)(\Theta_i')^2+m_i^2 a^2[1-\cos{\Theta_i}]} \nonumber \\ 
+ \frac{6H\Theta_i'\delta\Theta_i}{(\Theta_i')^2+2m_i^2 a^2[1-\cos{\Theta_i}]},
\label{ADMcovariant}
\end{eqnarray}
\begin{eqnarray}
\label{ADMvelpot}
\theta_{A,i} \equiv -2\frac{\delta\Theta_i}{\Theta_i'} .
\end{eqnarray}
For multi-component universes, we define an average covariant axion energy density as
\begin{eqnarray}
\Delta_{A,ave} \equiv \frac{\sum_{i}\epsilon_{A, i}\Delta_{A, i}}{\sum_{i} \epsilon_{A, i}}.
\label{eq:cov_2}
\end{eqnarray}
\subsubsection{Axion background energy}
\label{subsec:ABE}
After the first phase, the axion decay constant $f$, which determines the amplitude of background energy, must in principle be adjusted manually. In our computation, a method to automatically determine the decay constant has been designed. We describe how it is carried out by iterations below. The initial guess of the decay constant is determined from matching Hubble parameter at the matter-radiation equality exclusively with the radiation background. This decay constant, along with an $a^{-3}$ approximation for the axion background density, can be utilized in determining the Hubble parameter at the time long after the onset of mass oscillations and generating a modified decay constant. One can then iterate these steps to obtain a converged decay constant. 
\subsubsection{Schr\"{o}dinger equation}
\label{subsec:approx} 
To eliminate the rapid mass oscillation in the Klein-Gordon equation, Eq.~(\ref{apeom}), after the onset of mass oscillations, the Sch\"{o}dinger equation is adopted~\cite{Zhang1,Zhang2}. For the background angle, it can be approximated to be as $\Theta_i \approx \Theta_{0,i} a^{-3/2}\cos(\omega_{\Theta,i} t)$ with $\Theta_{0,i}$ being a constant and $\omega_{\Theta, i}^2 \equiv m_i^2 (1 - \Theta_{0,i}^2a^{-3}/8 )$. We introduce a new variable $\psi_i$ which relates to angle perturbation as $\delta \Theta_i \equiv \sqrt{\epsilon_{A, i}}\operatorname{Re}[\psi_i e^{-i\omega_{\Theta,i} t}]$, where $\operatorname{Re}$ denotes the real part. One can put these expressions back to Eq.~(\ref{apeom}) and ignore terms with frequency greater than or equal to $3\omega_{\Theta, i}$ and $O(H^2/m^2a^2)$ terms to derive the Schr\'{o}dinger equation. For a wave number $k$ mode perturbation, the resultant evolution equation is 
\begin{eqnarray}
i \omega_{\Theta, i}\dot{\psi_i} = \frac{1}{2}\left[ \left( \frac{k}{a} \right)^2 \psi_i-\alpha_i(\psi_i+\psi_i^*) \right] +m_i^2\phi, 
\end{eqnarray}
where $\alpha_i$ is defined as $\alpha_i \equiv \omega_{\Theta, i}^2 \Theta_{0,i}^2a^{-3}/8$ and the over dots denotes derivatives with respect to the cosmic time $t$. The $(\psi_i+\psi_i^*)$ term comes from the non-linear potential of the extreme axions, vanished for free axions.
\subsubsection{Matching condition}
\label{sec:mca}
The transition from the Klein-Gordon equation to the Schr\"{o}dinger equation requires the expression of covarinat energy density in relation to $\psi$. One can use the expression for $\Theta_i$ and $\delta \Theta_i$ from Sec.~\ref{subsec:approx} and apply the approximations stated in that subsection to derive the covariant energy density from Eq.~(\ref{ADMcovariant}):
\begin{eqnarray}
\Delta_{A,i} = 2 \psi_{R,i} - \frac{3H}{m_i a} \psi_{I, i},
\label{covariant_E}
\end{eqnarray}
where $\psi_R$ and $\psi_I$ are the real part and imaginary part of $\psi$, respectively. We stress that this post-transition expression of $\Delta_{A,i}$ given by the Schr\"{o}dinger equation does not possess mass oscillations. The first matching condition demands $\Delta_{A,i}$ to be continuous. The second matching condition is the derivative of Eq.~(\ref{covariant_E}) to be continuous. Both are to be used to solve for $\psi_R$ and $\psi_I$. Due to the tiny residual amplitude of mass oscillations of the covariant energy density, $\Delta_{A}$ is evaluated as the average between consecutive peak and trough immediately before the matching point. By the same token, the derivative of $\Delta_{A}$ must be represented by average slopes of consecutive peaks and troughs of$\Delta_{A}$.

Finding these peaks can be tricky, especially in two-component cases. As we switch the two components simultaneously to phase two, if the mass difference is wide, the slope of $\Delta_{A}$ for the massive component can sometimes be too large to detect peaks and troughs automatically. This is because no point with a zero slope is to be found numerically. Additionally, since the mass oscillations for the massive component can be too feeble to be defined at the transition point, it can be challenging to automatically determine the slope. We discuss more thoroughly the limitations of our scheme in Sec.~\ref{sec: PV} and Appendix~\ref{appd}.

Our choice of the transition time is at the end of the tenth mass oscillation cycle for the lightest component. The reason for choosing the lightest component as the benchmark is that it has the latest onset of mass oscillations.
\subsection{Photon and baryon}
\label{sec:pb}
\subsubsection{Friction equations}
\label{sec:fe}
The friction-coupled equations for photon and baryon perturbations are
\begin{eqnarray}
&&\delta_\gamma' = \frac{4}{3}k^2 \theta_\gamma + 4 \phi '  
\label{eq_delph} \\
&&\theta_\gamma' = -\frac{1}{4} \delta_\gamma - \alpha_T Q - \phi
\label{eq_thetaph} \\
&&\delta_b' = k^2 \theta_b + 3 \phi '  
\label{eq_delb}\\ 
&&\theta_b' + H \theta_b = \beta_T Q - \phi, 
\label{eq_thetab}
\end{eqnarray}
where $Q \equiv \theta_b - \theta_\gamma$, $\alpha_T \equiv a n_e \sigma_T$, where $n_e$ is the electron number density and $\sigma_T$ is the Thomson scattering cross section; $\beta_T \equiv (4\epsilon_\gamma/3\epsilon_b) \alpha_T$. The evolution of the neutrino perturbation behaves the same as the photon perturbation at the first half cycle of oscillation and damps abruptly after crossing the null~\cite{nuturnoff}. In our computation, we set the neutrino perturbation to be zero after crossing the null~\cite{Zhang1}.
\subsubsection{Diffusion approximation}
\label{sec:DA}
At the second phase, the diffusion approximation replaces the friction equations due to the strong Thomson scattering. The diffusion approximation works at the strong coupling limit ($k/\alpha_T\ll1$) and sub-horizon era ($k/H\gg1$). Under these circumstances, $Q$ should be small for the photon-baryon fluid. The $\phi '$ term will be ignored since it amounts to $O(H^2/k)$. Although the gravity term for the sub-horizon modes in radiation epoch should be overwhelmed by the radiation pressure, $\phi$ should be retained as it may slightly affect the damping and the photon oscillation pattern near the end of the second phase, especially for low-$k$ modes. With these assumptions, we can derive
\begin{eqnarray}
&&\delta_\gamma = -4\left[ Q'+(\alpha_T+\beta_T)Q-\frac{3H\delta_\gamma'}{4k^2} \right]
\label{delph}\\
&&\delta_\gamma'' = -\frac{k^2}{3}\delta_\gamma -\frac{4}{3}k^2 \alpha_T Q - \frac{4}{3}k^2\phi 
\label{dd_delph}
\end{eqnarray}
from Eq.~(\ref{eq_delph}), Eq.~(\ref{eq_thetaph}) and Eq.~(\ref{eq_delb}). To eliminate the $Q$ terms, Eq.~(\ref{delph}) and Eq.~(\ref{dd_delph}) can be combined into 
\begin{eqnarray}
\delta_\gamma'' + \frac{k^2}{3}\left( \frac{\beta_T}{\alpha_T+\beta_T} \right)&&\delta_\gamma =\frac{4}{3}k^2 \left(\frac{\alpha_T}{\alpha_T+\beta_T}\right)Q'
\nonumber\\
-&&\left( \frac{\alpha_T}{\alpha_T+\beta_T}\right) H\delta_\gamma'-\frac{4}{3}k^2\phi.
\label{dddeleq}
\end{eqnarray}
For a relatively small $Q'$ and $\phi$, the leading order oscillation frequency for $\delta_\gamma$ can be approximated to be $\sqrt{k^2\beta_T/3(\alpha_T+\beta_T)}$. With Eq.~(\ref{eq_thetab}), the term for $Q'$ can be rewritten as below,
\begin{eqnarray}
\frac{4}{3}k^2 \left(\frac{\alpha_T}{\alpha_T+\beta_T}\right)Q' = -\left[\frac{k^2\alpha_T}{3(\alpha_T+\beta_T)^2}\right] \delta_\gamma'.
\label{Qprime}
\end{eqnarray}
Finally, the diffusion equation for the photons becomes
\begin{eqnarray}
\delta_\gamma'' + \left(\frac{\alpha_T}{\alpha_T+\beta_T}\right)&& \left[ \frac{k^2}{3(\alpha_T+\beta_T)}+H\right] \delta_\gamma' \nonumber \\
&&+ \frac{k^2\beta_T}{3(\alpha_T+\beta_T)} \delta_\gamma = -\frac{4}{3}k^2\phi.
\label{diffusion_ph}
\end{eqnarray}
For the baryon density perturbation, one can combine Eq.~(\ref{eq_delph}), Eq.~(\ref{eq_delb}) and Eq.~(\ref{eq_thetab}) into 
\begin{eqnarray}
\delta_b'' = k^2 \beta_T Q - \frac{3}{4}H\delta_\gamma' - k^2 \phi,
\label{dddel_b}
\end{eqnarray}
and eliminate the $Q$ term via Eq.~(\ref{delph}), and then replace the $Q'$ term with Eq.~(\ref{Qprime}). The final result is
\begin{eqnarray}
\delta_b'' = -\frac{k^2 \beta_T}{4(\alpha_T+\beta_T)^2}\delta_\gamma' -&&\frac{3\alpha_T H}{4(\alpha_T+\beta_T)}\delta_\gamma' \nonumber \\
&& -\frac{k^2 \beta_T}{4(\alpha_T+\beta_T)}\delta_\gamma -k^2 \phi.
\label{diffusion_b}
\end{eqnarray}
Eq.~(\ref{diffusion_ph}) and Eq.~(\ref{diffusion_b}) are the equations with the diffusion approximation.

In the third phase, a set of friction-coupled equations resumes. Since the photon-baryon drag becomes loosely coupled at this stage, $Q$ is finite. Thus, instead of evolving $\delta_\gamma$, $\delta_b$, $\theta_\gamma$, and $\theta_b$ as the standard friction equations do, we choose to evolve $\delta_\gamma$, $\delta_b$, $\theta_\gamma$, and $Q$ while $\theta_b$ can be retrieved via a direct addition. The evolution equation for $Q$ is the subtraction between Eq.~(\ref{eq_thetaph}) and Eq.~(\ref{eq_thetab}). The method to obtain an accurate initial condition of $Q$ will be discussed in the next subsection.
\subsubsection{Matching condition}
\label{subsec:mcbr}
From the second phase to the third phase, there will be a sizable numerical error if one derives the matching condition for $Q$ via direct subtraction between $\theta_\gamma$ and $\theta_b$ since both photon and baryon velocity potential are large quantities whereas $Q$ itself is small. In light of this problem, a better approach is evaluating the matching when the photon oscillation is at its peak, where Eq.~(\ref{delph}) can be rewritten as
\begin{eqnarray}
Q' +  (\alpha_T+\beta_T+H) Q = -\frac{\delta_\gamma}{4} .
\label{Qprime_der}
\end{eqnarray}
The previously ignored $HQ$ has been brought back here. Due to the aforementioned oscillation frequency, the $Q'$ term is of order $O(kQ)$, which is much smaller than other terms. Therefore, by ignoring the $Q'$ term, the matching condition is
\begin{eqnarray}
Q = -\frac{\delta_\gamma}{4(\alpha_T+\beta_T+H)}.
\label{IC_Q}
\end{eqnarray}

From phase one to phase two, we choose the transition criteria to be $k/H=10$ if this transition is later than the tenth mass oscillation; otherwise, we use the end of the tenth mass oscillation as the transition point. From phase two to phase three, we choose $k/\alpha_T = 0.1$ as our transition criterion.

\section{Single component Universe}
\label{sec:onecom}
The validity of the approximation scheme is presented in this section. In the following discussions, the scale factor at the matter-radiation equality will be represented by $a_{eq}$. All axion masses are chosen to be $10^{-22}$eV in this section. 

\subsection{Performance and validity}
\label{sec: PV}
\begin{figure}[h]
\includegraphics[width=8.6cm]{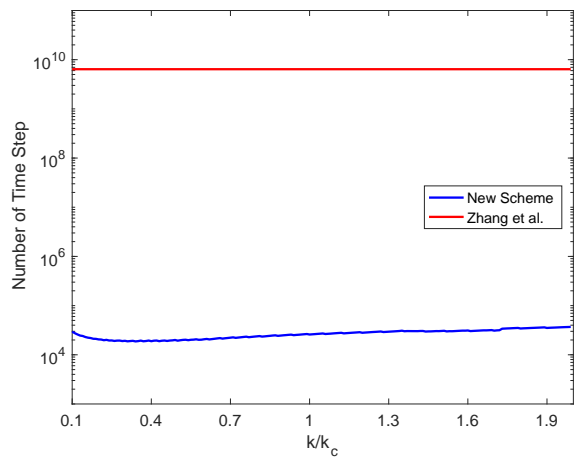}
\caption{\label{timestep}Numbers of computation steps for the approximation scheme and the full evolution scheme of Zhang et al~\cite{Zhang1,Zhang2}. The integration interval starts from $ 4\times 10^{-5}a_{eq}$ to $a_{eq}$ for both schemes, where $a_{eq}$ is the scale factor of the matter-radiation equality. The critical wave number $k_c \approx 1/122$ kpc$^{-1}$.}
\end{figure}
Comparisons of the approximation scheme with the full evolution are shown in this subsection. First, the number of time steps has been greatly reduced with our approximations. Fig.~\ref{timestep} demonstrates the efficiency of time step reductions. Here $k_c$ is the comving Compton wavenumber when $m = 2H/a$, for which $k_c/a = m = 2H/a$ ($k_c \approx 1/122$ kpc$^{-1}$). The number of integration variables is $6$ for both approximate and exact methods. With integration interval from $z\sim10^6$ to $z\sim10^3$, it reveals a reduction in time step number from $O(10^{10})$ to $O(10^{4\sim5})$. It should be noted that the time steps for our computation are different for different wavelengths. Incurring numerous matter-wave oscillations for short-wavelength modes, it would entail higher resolution. On the other hand, more time steps are required for long-wavelength modes since the integration switches to the approximation scheme only after the mode entered the horizon. The lower the $k$ is, the later the transition happens, and hence more integration steps.

\begin{figure}[htbp]
\includegraphics[width=8.6cm]{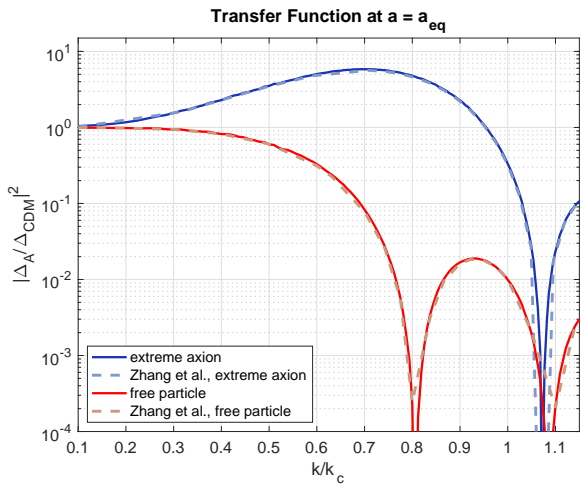}
\caption{\label{1compare}  Good agreement of transfer functions of our approximation scheme (solid lines) and of the full evolution scheme (dashed lines) by Zhang et al. evaluated at $a_{eq}$. The initial axion angles of the background field are nearly $90^{\circ}$ for the free particle cases and $179.8^{\circ}$ for the extreme axion cases.}
\end{figure}

The transfer functions comparing our scheme and the full-evolution scheme is presented in Fig.~\ref{1compare}. A free particle case and an extreme axion case are both compared in this figure. The background axion's initial angle is $90^{\circ}$ and $179.8^{\circ}$ for the free particle scenario and extreme axion scenario, respectively. Good agreement between the two schemes is found. The discrepancy between our scheme and the full evolution scheme is at most $5\%$ except the regions near the nulls of matter-wave oscillations, where errors are primarily affected by the resolution in Fig.~\ref{1compare}. 

There are two limitations of our scheme. First, the matching point from phase one to phase two must be respectably earlier than the matter-radiation equality. As the mass oscillations get closer to the matter-radiation equality, the non-linear effect of the background field starts to contribute, and the previously mentioned $a^{-3}$ extrapolation is not accurate. Since the onset of mass oscillations for axion mass $10^{-26}$eV is about the same time of matter-radiation equality, the approximation adopted in this work is impracticable with particle mass lower than this value. Empirically, our approximations work the best with an axion mass larger than $10^{-25}$eV. 

Another limitation is that our computation fails to function at extremely low-$k$ modes. Our experience tells that a numerical instability in our computation for $k/k_c$ smaller than $0.02$. The primary reason for this is the late transition for very low-$k$ modes. Since field matching is well after the onset of mass oscillations and the mass oscillations in $\Delta_A$ have been greatly erased, it is difficult to detect the mass oscillation peaks used to determine the matching conditions for the Schr\"{o}dinger equation. Another reason derives from the full-evolution $\phi$. With a replacement of the passive-evolution $\phi$, this problem can be solved for $k>0.01k_c$. For longer wavelengths, an accurate description for $\phi$ is necessary~\cite{Zhang1}. This long-wave issue will be discussed more thoroughly in Appendix~\ref{appd}. This problem will be most prominent in two-component universes, and the pertinent discussion is given in the next section.

There is a trade-off between the accuracy and performance of our scheme. The error can be improved to less than $1\%$ but with a later transition point; that is, more time steps are required. Additionally, if one chooses a late transition point, the range for viable $k/k_c$ and masses will be reduced due to the previously stated reasons. 
\subsection{Baryon acoustic oscillations and velocity potential}
\label{subsec:bao}
\begin{figure}[htbp]
\includegraphics[width=8.6cm]{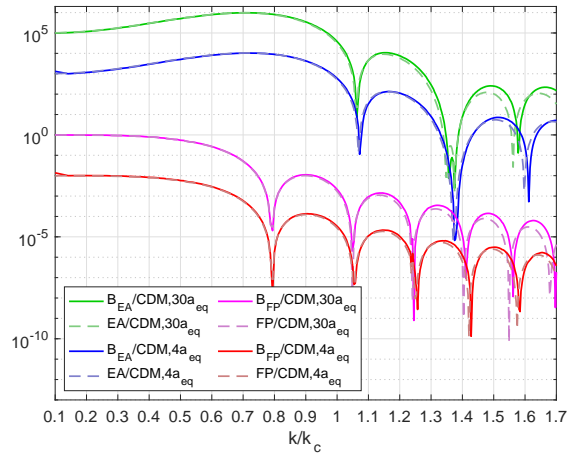}
\caption{\label{baryon} Comparison of baryon (solid lines) and dark matter (dashed lines) transfer function particularly on the slight phase shifts of matter wave oscillations. Here ``FP" and ``EA" are the abbreviations for ``free particle" with an initial axion angle near $90^{\circ}$ and ``extreme axion" with an initial axion angle at $179.96^{\circ}$, respectively. ``B" denotes ``baryon". The ``$j$/CDM" means $|\Delta_j/\Delta_{CDM}|^2$ for the $j$ component. It should be noted that the curves for baryons are, in fact, roughly two orders of magnitude and $0.7$ lower than those for axions at $4a_{eq}$ and $30 a_{eq}$, respectively, but we scale them to the same amplitude for the low-$k$ modes to optimize the visualization. In addition, all curves at very low $k$ are near $1$, but we multiply these curves by $10^5$, $10^3$, $1$, $10^{-2}$ for better visualization.}
\end{figure}

The baryon to the axion phase shift is shown in Fig.~\ref{baryon}. The decreasing electron as a result of the increasing recombination has been accounted for via the Saha equation~\cite{Saha}. In this figure, $4a_{eq}$ is chosen to show the phase difference near the epoch of recombination while $30a_{eq}$ ($z\sim100$) is chosen for initial conditions of simulations. This small-scale phase shift is due to the comparable Hubble friction that makes baryon perturbation react late to the gravity. The phase aligns well at the recombination epoch since the ADM gravity dominates at all shown scales. 

\begin{figure}[htbp]
\includegraphics[width=8.6cm]{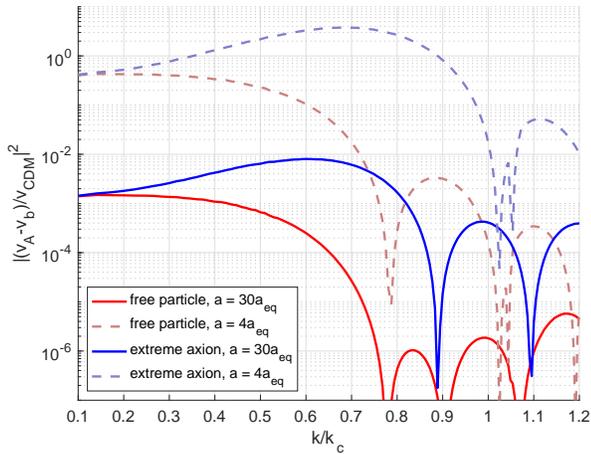}
\caption{\label{veldiff} Axion-baryon relative velocity at $4a_{eq}$ and $30 a_{eq}$. It can be seen that the velocity difference is large immediately after the recombination comparing to that in the late time ($z\sim100$).}
\end{figure}
It has been suggested that considerable large scale velocity difference between dark matter and baryons at the early epoch would significantly affect structure formation nonlinearly~\cite{dmbaryonflow}. This substantially smaller baryon velocity from that of the dark matter at the recombination epoch is due to the severe photon-baryon drag from which dark matter does not suffer. We present such velocity difference in Fig.~\ref{veldiff}. It can be seen that the relative velocity becomes negligible at $z\sim100$ because the enhanced dark matter gravity pulls the baryons to align with the dark matter. This phenomenon may imply that simulations starting from late matter epoch are unable to subsume the nonlinear effects originated from the axion-baryon relative velocity.

\begin{figure}[htbp]
\includegraphics[width=8.6cm]{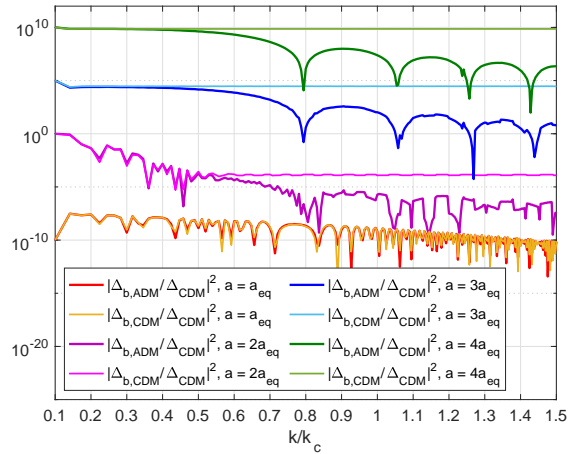}
\caption{\label{baryon_evo} Comparisons of small scale acoustic oscillations of baryons for ADM and CDM cases. All the lowest $k$ modes of these curves are scaled to $1$, and then the curves at $a_{eq}$, $2a_{eq}$, $3a_{eq}$, and $4a_{eq}$ are multiplied by overall factors $10^{-10}$, $1$, $10^5$, and $10^{10}$, respectively. This plot shows baryon acoustic oscillations for ADM and CDM cases are indistinguishable at low $k$ and before $a_{eq}$, but different at moderate $k$ and after $a_{eq}$. Moreover, baryon acoustic oscillations get erased after $3a_{eq}$ for the range of $k$ ($\sim 10$ Mpc$^{-1}$) of interest in this paper.}
\end{figure}

The small-scale baryon acoustic oscillations are difficult to observe but interesting to explore. They differ in CDM and ADM models, as revealed in Fig.~\ref{baryon_evo}. Before the recombination at $a\sim 3 a_{eq}$, baryon acoustic oscillations in the ADM universe was indistinguishable from those in the CDM universe. However, after the recombination, the baryon perturbation in the CDM case departs from the ADM counterparts  since the baryon perturbations are pulled by the gravity of different dark matter perturbations, where the CDM perturbation grows without pressure supports while the ADM perturbation is suppressed by the quantum pressure at sufficiently high $k$. 

\begin{figure}[htbp]
\includegraphics[width=8.6cm]{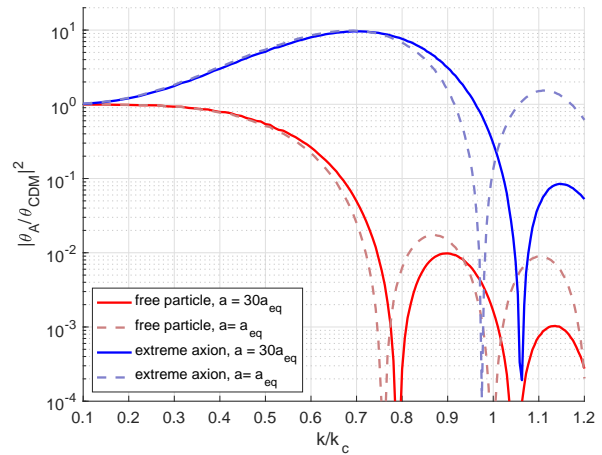}
\caption{\label{velocity} Transfer functions of velocity potentials for free particles and extreme axions at $a_{eq}$ and $30a_{eq}$. Again, the axion angle is $179.96^{\circ}$ for extreme axion cases. We note that the velocity transfer function has a spectral bump immediately before the cutoff, similar to the density transfer function. }
\end{figure}
In the past, velocity potential for fuzzy dark matter simulations has prevalently been set similar to cold dark matter~\cite{velocity}. This can be problematic for the modes shorter than the spectral cutoff. We are able to compute the velocity potential free from the cold dark matter assumption and provide an accurate description of axion velocity near the spectral cutoff. Such a velocity potential is given in Fig.~\ref{velocity}. It reveals slightly different behaviors at the epoch of matter-radiation equality and at the time when simulations usually start ($z=100$). One can see that the velocity transfer functions share similar features as the density transfer functions. That is, the amplitude of the density perturbations correlates well with that of velocity perturbations.

\section{two-component axion universe}
\label{sec:twocom}
\begin{figure*}[htbp]
\includegraphics[width=8.6cm]{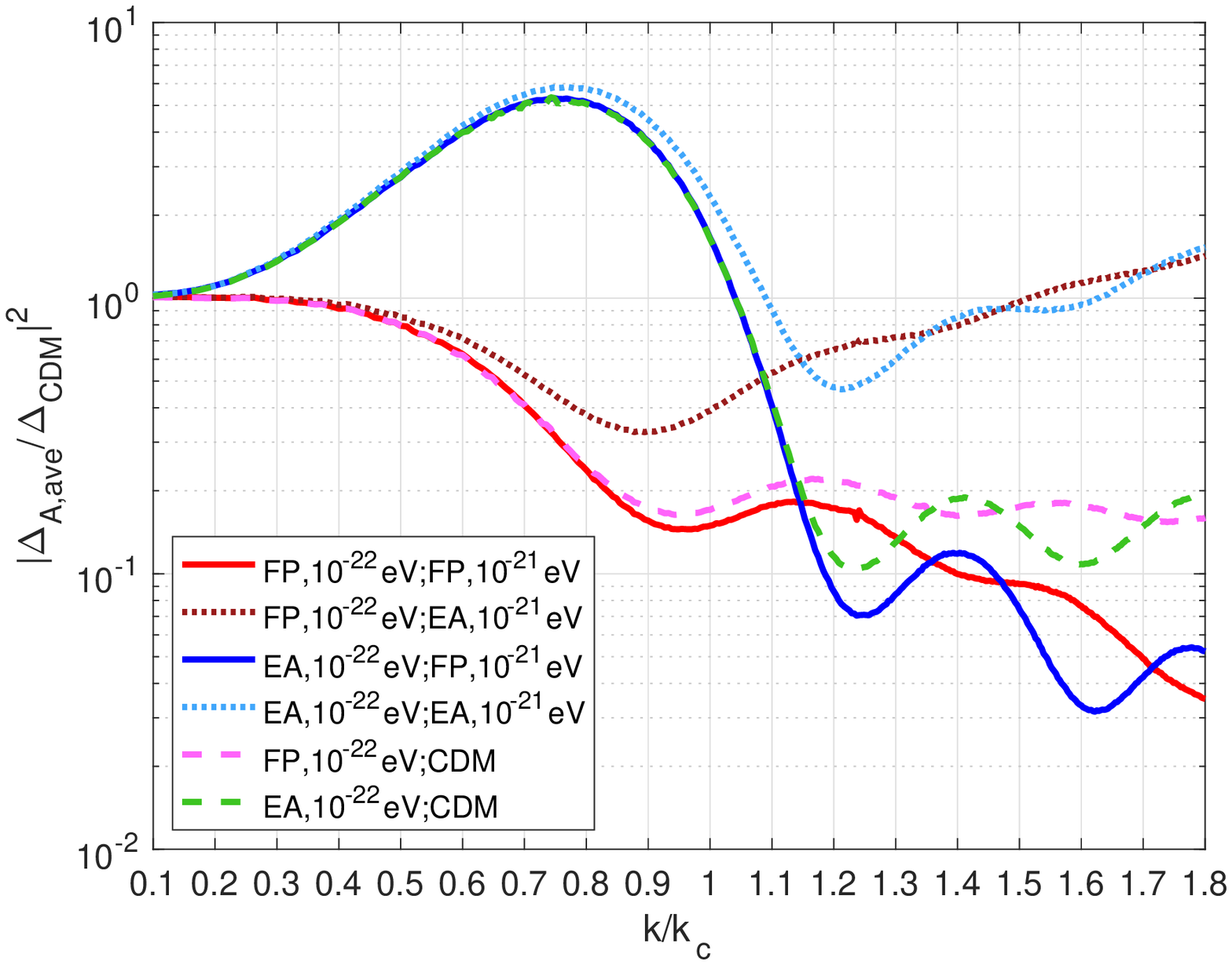}
\quad
\includegraphics[width=8.6cm]{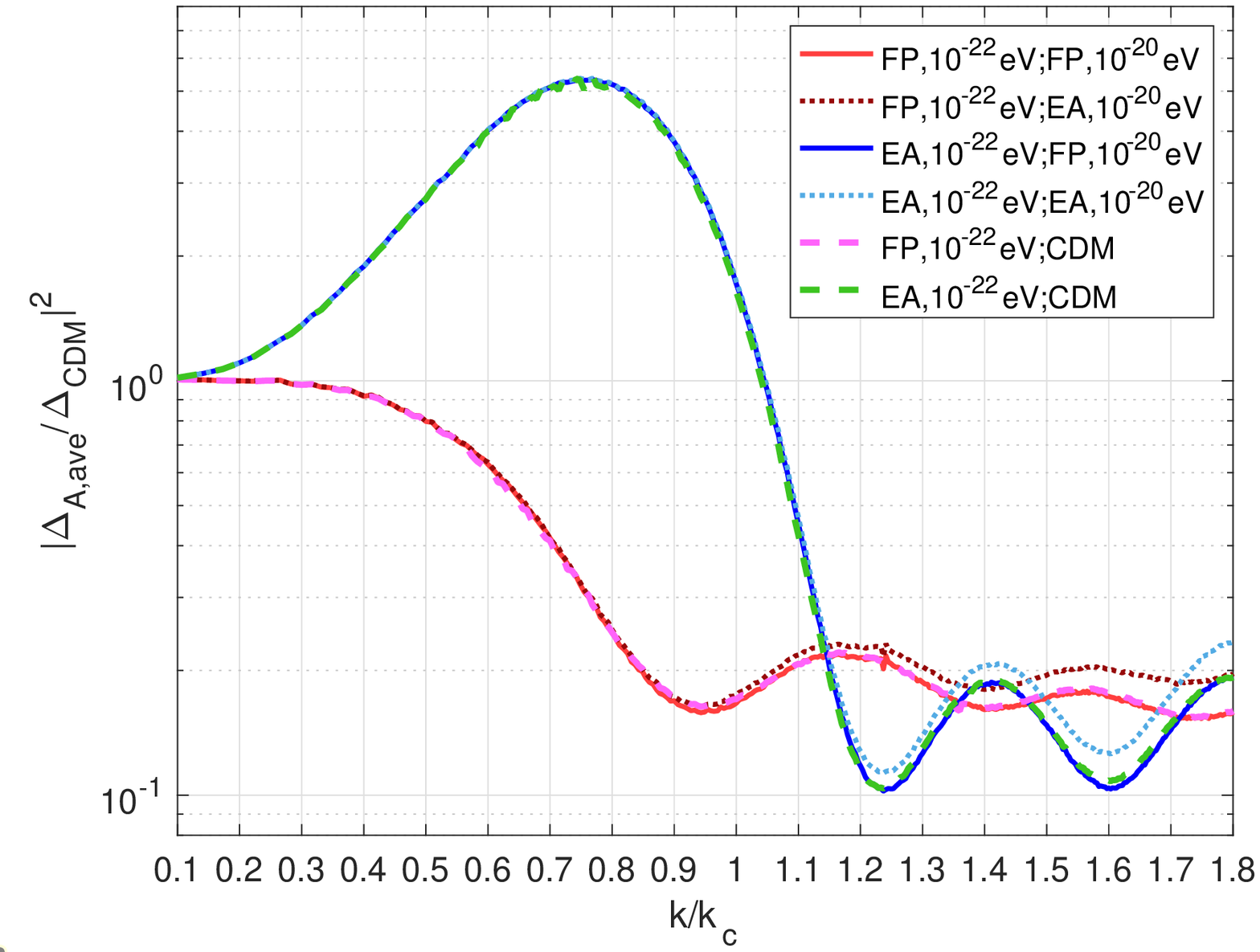}
\caption{\label{fig:twocomp}Transfer functions at $a_{eq}$ of two-component universes with various compositions. The left panel illustrates cases with a smaller axion mass difference where the two components consist of $10^{-21}$eV and $10^{-22}$eV axions, while the right one illustrates cases for a larger axion mass difference composed of $10^{-20}$eV and $10^{-22}$eV axions. The ``ADM/CDM" configuration is presented in both panels as references, particularly showing $10^{-20}$eV axions can be replaced by CDM to good precision. The data points from $0.1k_c$ to $0.33k_c$ for the ``$10^{-20}$eV/$10^{-22}$eV" cases are replaced with the ``ADM/CDM" universe on the right panel.}
\end{figure*}

Various configurations of the two-component axion universe are investigated in this section. The background energies of two kinds of dark matter are chosen to be equal for all cases unless specified otherwise. All approximations in Sec.~\ref{sec:Approx} can be directly applied to calculations in this section. Light component serves as the benchmark to determine the first transition points since they start their mass oscillations later than massive ones. The critical $k$ for $10^{-22}$eV axions is denoted as $k_c$ in this section.

Transfer functions at the matter-radiation equality are presented in Fig.~\ref{fig:twocomp}. The average covariant energy density of ADM is defined in Eq.~(\ref{eq:cov_2}). The average spectra can grossly be understood as the superposition of features in individual components. For example, the small-scale spectral bump is due to the bump feature of the massive extreme axions. For the case with a sizable axion mass difference, the energy density of the lighter component is severely suppressed at the bump scale. Whether the resulting spectral power would exceed the power of CDM actually depends on mass differences and chosen initial angles. It should be noted that since the bump feature for extreme axions only affects a certain scale, the high-$k$ bump feature does not show up for $10^{-20}$eV extreme axion in Fig.~\ref{fig:twocomp}. Nevertheless, it cannot be over-emphasized that the bump feature of this case always emerges at sufficiently large wave numbers. 

The small scale bump could be a new resolution to the ADM Lyman-$\alpha$ issue. Much like the extreme axion solution proposed in~\cite{lyman-a}, the two-component universe can also provide a strong power at the small scale to solve the Lyman-$\alpha$ problem. Moreover, the two-component model offers more degrees of freedom. While the initial angle is the only degree of freedom in the one-component axion universe, the additional mass and density ratio parameters are to make the resulting spectra more flexible. 

In addition, the spectral excess for light extreme axions is preserved as well. The massive component only affects the scale of its spectral cutoff and bump. For the axion masses under consideration, the massive component resembles CDM at the scale where the light component discloses its spectral features; thus, the large-scale feature is determined by the light component exclusively.

For the $10^{-22}$eV/$10^{-20}$eV cases, the range $k<0.33k_c$ has been replaced by the ADM/CDM case, because the calculation for the massive component is highly unstable around $0.1k_c$ as indicated in Sec.~\ref{sec: PV}. The spectral computation for even larger mass differences is far from practical. We offer an alternative here. Since the large-scale character of axion dark matter approaches that of the cold dark matter, we can replace very massive axions by cold dark matter. However, replacing massive ADM components by CDM can only apply to free particle axions. CDM replacement can never capture the bump feature arising from extreme axions. 

There are two options if one wants to compute the spectrum when the massive component is an extreme axion. First, one can exploit the approximation of the Schr\"{o}dinger equation at super-horizon. For that, one should add the derivative of metric perturbation as the additional source term. Although this method may capture the overall feature of the power spectrum, it may create errors up to $10\%$ at long wavelengths. The other method is to evaluate the power spectrum only for $k>0.02k_{c, massive}$, with $k_{c, massive}$ being critical $k$ of the massive components while utilizing the CDM/ADM construction for even larger scale. 

\begin{figure}[htbp]
\includegraphics[width=8.6cm]{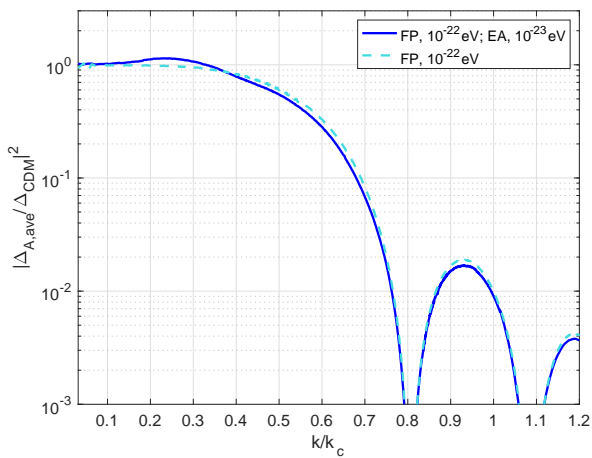}
\caption{\label{TF2223} Transfer functions of the two-component universe at $a=a_{eq}$ for a major component with $10^{-22}$eV free particles and a minor component with $10^{-23}$eV extreme axions where the density ratio is $29:1$. The dashed line is the one-component free particle case with axion mass $10^{-22}$eV for comparison. It shows that a noticeable low-$k$ spectral bump can be generated by a tiny fraction of the second light extreme axion component.}
\end{figure}
We finally present a case that could be responsible for the existence of very high-redshift quasars~\cite{massgal1,massgal2}. The existence of these quasars normally implies the formation of hosting massive galaxies at redshift $z>10$~\cite{z11galaxy}. To achieve this, strong power at the large scale is necessary. Fig.~\ref{TF2223} shows a case with background densities $29:1$ for $10^{-22}$eV free particles to $10^{-23}$eV extreme axions. A small amplitude bump of $15\%$ at the large scale is attributed to the light extreme axion. This spectrum provides a possibility for seeding early massive host galaxies for supermassive black holes. The mass of the collapsed halo is estimated to be $(4\pi/3) (\lambda/2)^3 \rho_0 \sim$ O($10^{11\sim12}$)$\textup{M}_\odot$, where $\lambda$ is the comoving wavelength ($2\pi/k$) and $\rho_0$ is the present background density~\cite{halomass}. In this case, $k$ is taken to be $0.2 \sim 0.3 k_c.$ This choice is only a particular example. It has sufficient degrees of freedom to alter axion masses, axion initial angles, and background energy ratios to obtain other results with different galaxy masses and galaxy formation times. The axion masses determine the galaxy mass, whereas the initial axion angles and the background energy ratio could affect the galaxy formation time.
\section{Conclusions}
\label{sec:conclusion}
We present an efficient scheme for computing perturbations in the axionic universes. The approach invokes approximation exploiting the Schr\"{o}dinger equation and the diffusion approximation. With these approximations, the reduction by six orders of magnitude in computational time is achieved, while the accuracy is higher than $95\%$. Our computational code is available on GitHub\footnote{See https://github.com/YiHsiungHsu/MASTER.}

Our computation demonstrates the novel spectra in two-component universes. A massive extreme axion component can enhance the power at small scales without changing the large-scale spectrum. This scenario offers a new perspective to compare with the Lyman-$\alpha$ observations. Moreover, the incorporation of a massive free particle does not affect the spectral cutoff substantially. That is, the large scale structures and missing dwarf galaxies may both be solved by the two-component model. Two-component models offer a possibility of coexisting solitons of different sizes at the centers of galaxies~\cite{solitoninsoliton}. The soliton within another soliton scenario may solve the age problem of the star cluster at the center of the dwarf galaxy (Eridanus II)~\cite{Li_2017, starcluster}. However, due to the non-linearity nature of the soliton, further simulations for this scenario are needed. For the first time, our calculations can offer the initial conditions for densities and velocities and would greatly improve future ADM simulations.

Finally, the puzzle for the existence of the massive quasar hosting galaxies at $z\geq10$ could be resolved via our two-component universes. We illustrate a particular case where one major component is free particles with axion mass $10^{-22}$eV, and the other minor component is extreme axions with axion mass $10^{-23}$eV. The resulting large scale spectral bump from the minor light component corresponds to the halo mass of O($10^{11\sim12}$) $\textup{M}_\odot$ that yields the first-generation galaxies.

\begin{acknowledgments}
This research is supported by the Ministry of Science and Technology (MOST) of Taiwan under Grants No. MOST 107-2119-M-002-036-MY3, and the NTU Core Consortium project under Grants No. NTU-CC- 108L893401 and No. NTU-CC-108L893402.
\end{acknowledgments}

\appendix
\section{Long-wave issue}
\label{appd}
In this section, we discuss the origin of, and the solution for, the long-wave problem. It has been mentioned that as $k \leq 0.02 k_{c22}$, the computation is prone to fail, where $k_{c22}$ is the critical $k$ of $10^{-22}$eV ADM. The first issue is the inaccurate determination of rapidly decreasing mass oscillations in $\Delta_A$ long after the onset of mass oscillations. To solve this problem, the derivative is approximated by the slope of a distant section in $\Delta_A$ immediately before the transition. However, the accompanied error may go up to $10\%$. On the other hand, this method only applies to $k\leq0.01k_{c22}$. For an even smaller $k$, the mode enters horizon near the matter-radiation equality, and these modes are beyond the scope of this work.

Another issue is subtraction in the denominator of metric perturbation:
\begin{widetext}
\begin{equation}
\phi = \frac{4 \pi G \{a^2[\epsilon_\gamma \Delta_\gamma +\epsilon_b \Delta_b +\epsilon_\nu \Delta_\nu]+ \sum_{i}[
\Theta_i'\delta\theta_i'+m_i^2a^2\sin{\Theta_i}\delta\theta_i +3H\Theta_i'\delta\theta_i]\}}{-k^2+4 \pi G \sum_{i}(\Theta_i')^2}.
 \label{potential}
\end{equation}
\end{widetext} 
In most cases, the $k^2$ term dominates the denominator. However, as the wavelength increases, the $4\pi G(\Theta')^2$ term would surpass the $k^2$ term. There will be a zero-crossing in the denominator, but the numerator should also be zero at the crossing. Hence, the error of $\phi$ can be substantially amplified. Our suggested solution is replacing the full-evolution $\phi$ (Eq.~(\ref{potential})) with the passive evolution $\phi$ at an early time. The passive evolution metric perturbation should be similar to the full-evolution counterpart at early epochs as a result of the radiation dominance. The passive evolution potential has an analytical expression~\cite{Zhang1}, which can eschew the zero-crossing. This solution can only be valid for modes entering the horizon before the matter-radiation equality, which is the $k$ range we consider in this work.

Another solution to both problems is utilizing the super-horizon Schr\"{o}dinger equation after the onset of mass oscillations. As mentioned at Sec.~\ref{sec:twocom}, this equation requires an additional term involving $\phi'$. The first problem, the late transition problem, can be unraveled since the transition time is sufficiently early. On the other hand, the expression of potential after the transition to the Schr\"{o}dinger equation invokes no singular denominator. It is the sum of all covariant energy density, and hence, the zero-crossing conundrum can be avoided. However, as previously mentioned, an error up to $10\%$ would be accompanied by this alternative. 

\bibliography{paper.bib}

\end{document}